# An Analysis of the Prevalence of Pneumonia for Children under 12 Year Old in Tawau General Hospital, Malaysia

Suliadi F. Sufahani[1], Siti N. A. Mohd Razali[2], Mohammad F. Mormin[3], Azme Khamis[4]

[1,2,4]Departent Of Mathematic and Sciences, Faculty of Science, Art and Human Development, University Tun Hussein Onn Malaysia, Parit Raja, Batu Pahat, Johor, Malaysia

*suliadi@uthm.edu.my  asyikinr@uthm.edu.my azme@uthm.edu.my*

[3]Hospital Besar Tawau, Peti Surat 67, 91007 Tawau, Sabah, Malaysia

*antiarekey@yahoo.co.uk*

April 2012

Pneumonia is one of the serious illnesses, which involves lung infection specifically alveoli. Nearly 40,000 to 70,000 people die each year in United State because of pneumonia. Therefore, it is not a surprise that pneumonia is one of the most critical illnesses for children under 12 years old in many parts of the world, including Malaysia and particularly in Tawau, Sabah, Malaysia. The objectives of this study are: to develop a summary on the prevalence of pneumonia in Tawau General Hospital, to analyze the best practice to prevent this illness and lastly to determine an overview of which area that is widely affected by pneumonia. The results can assist doctors and the government to take major precautions and preventive measures efficiently to the full extent. This paper presents a descriptive analysis of the data, which are retrieved from the medical reports at the Tawau General Hospital. Through the findings, pneumonia is widely spread among young children under 12 years old. There are more than one major factor that leads to this critical illness, such as family background, genetic and environment. Therefore, the government, doctors and parents should take major steps to prevent children suffering from pneumonia.

## 1. INTRODUCTION

Pneumonia is one of the serious illnesses, which involves lung infection specifically alveoli. Nearly 40,000 to 70,000 people die each year in United State because of pneumonia. In Europe and North America, it was found that the prevalence of pneumonia for children under 5 years of age is 34 to 40 cases per 1000, which is higher than in any other time [1, 2, 3, 4, 5]. Therefore, it is not a surprise that pneumonia is one of the most critical illnesses for children under 12 years old in many parts of the world, including Malaysia and particularly in Tawau, Sabah.

The definition of pneumonia is too subjective. However, according to [3], pneumonia is defined as the presence of fever, acute respiratory symptoms, or both, plus evidence of parenchymal infiltration on chest radiography. In the worse situation, it can affect our lung where the alveoli are filled with pus and fluid, which makes breathing painful and limits oxygen intake. There are a large number of microorganisms and viruses that can cause pneumonia, such as respiratory syncytial virus, influenza virus, parainfluenza viruses, and adenovirus [1, 2, 3, 4, 5]. These viruses and microorganisms can be found in a child's nose or throat, and can be spread via air-borne droplets from a cough or sneeze and blood during afterbirth. The symptoms of pneumonia can be seen through the difficulty of breathing, continuous coughing, frequent fever, chills, loss of appetite, and wheezing.

Serious preventions have to be taken by parents, government and medical centre in order to avoid more pneumonia phenomena among children, such as taking immunization against Hepatitis B, having adequate intake of nutrition, providing affordable and clean indoor stoves and encouraging good hygiene in crowded homes. In addition, continuous research may assist them to view the latest statistics so that any prevention and action plan can be done to control the infection of pneumonia among children.

This paper is organized as follows. Section 2 presents the related data of pneumonia taken from Tawau General Hospital. Section 3 describes research methodology of analyzing the data. Section 4 shows the results and analysis obtained and Section 5 concludes the paper.

## 2. DATA

This paper presents an analysis of the secondary data that were obtained from the Tawau General Hospital. A patient, who comes to hospital, will be asked to fill out a special form before the actual medical check-up. The form requires some information such as the patient age, area of origin, parent's smoking background, parent's medical background (if known), patient medical background (if known) etc. Recent research has found that analytical research relies heavily on secondary data in terms of model construction and estimation. This study analyzed the patient profile data that were collected for a 6-month period, from June 2010 to December 2010. Due to highly confidential data, a total number of 102 patients' profile data were allowed to be shared. Therefore, the findings of this study were based on a sample size of 102 patients.

Figure 1 shows the number of patients who visited Tawau General Hospital from June 2010 to December 2010. The data show the number of patients who visited the Tawau General Hospital because of pneumonia. The diagram shows the highest number of patients who visited the hospital was in October. This is because October is categorized as part of the monsoon season. The weather is cold and clammy; therefore it could also lead to an increment of children suffering from pneumonia.

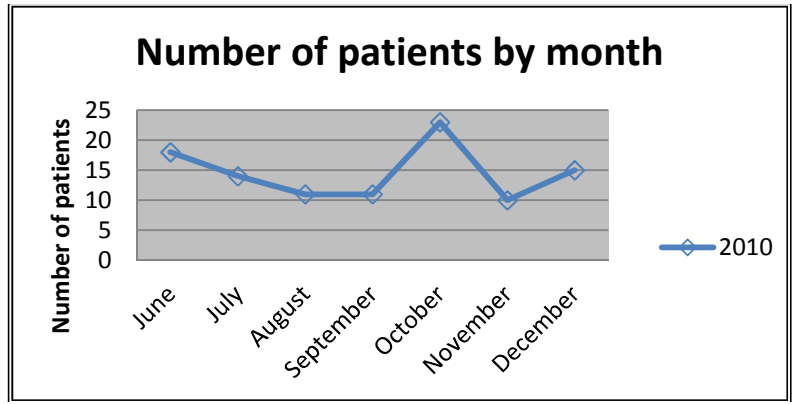

Figure 1: Number of patients who visited Tawau General Hospital from June 2010 to December 2010

### 3. RESEARCH METHODOLOGY

The data comprising 102 patients were analyzed by using a statistical quantitative method along with Microsoft Excel. We started with the descriptive analysis, where we wanted to see the overall pattern of the data set. The statistical techniques normally involved in the descriptive analysis are percentage, frequency distribution, mean, median, variance, standard deviation and coefficient of variation. Frequency distribution is a multiple-column table containing information on certain parameters; it is presented in frequencies and percentages. Mean and median are the measurements of average values on all data cases. Variance, standard deviation and coefficient of variation are the measurements of variability values of a variable. They are known to be measures of the reliability of the mean. Most researches used these techniques at the initial stage of data analysis in order to explore the data and understand their characteristics with their statistical implications [1].

### 4. RESULTS AND ANALYSIS

This section describes the data analysis and the results obtained from the study. The main objective of this study is to develop a summary profile of the patients who came to Tawau General Hospital because of pneumonia. This summary is graphically shown in Figure 2 to Figure 7. From the findings, Figure 2 shows that children aged 2-3 years old are mostly infected by pneumonia compared to other age categories. They are still considered as infants. Based on the sample of the patients, the number of percentage of parents who smoked is 37.25 percents and parents who did not smoke are 62.75 percents. This is shown in Figure 3. However, we only consider a small sample of the population. This sample of 102 patients yields a small portion of results on the smoking percentage.

Based on Figure 4, we can see that most of the patients were admitted for 3 to 4 days. There was a case where a patient was admitted for 14 days. Probably it involves a serious case of pneumonia. Figure 5 shows that 15.69 percents of the parents were infected by pneumonia since childhood. Therefore, the children might have inherited the illness from their parents.

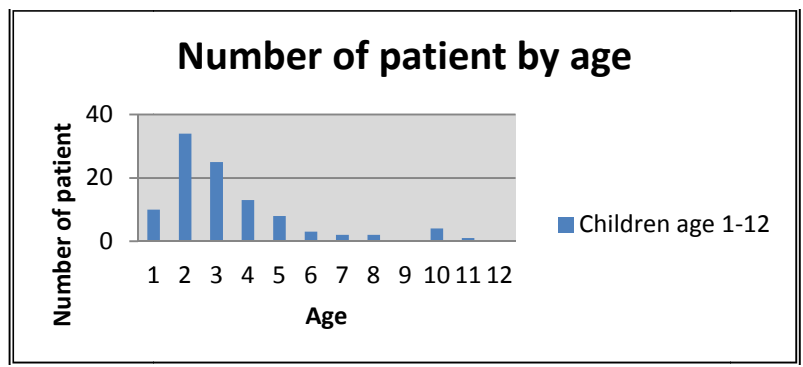

Figure 2: Number of patients based on age category.

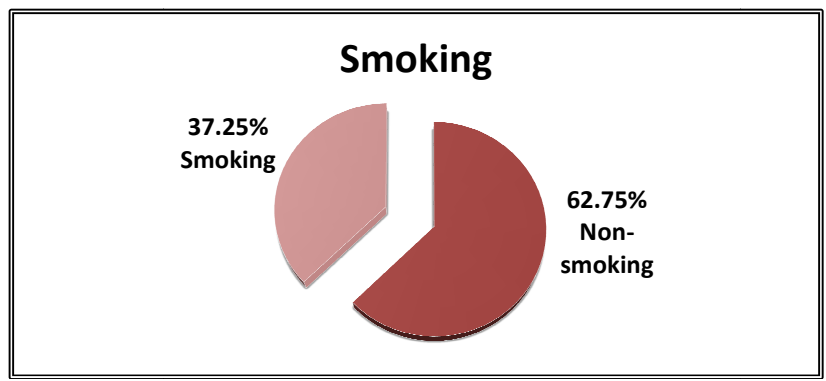

Figure 3: Percentage of parents who smoke.

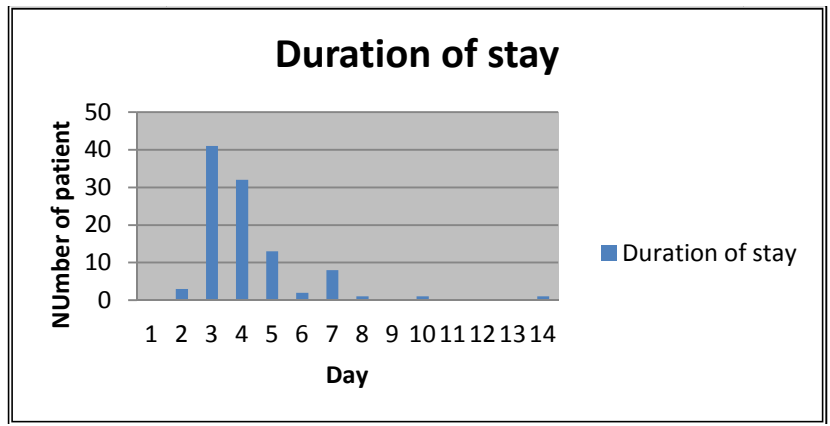

Figure 4: Number of patients admitted in the hospital by days.

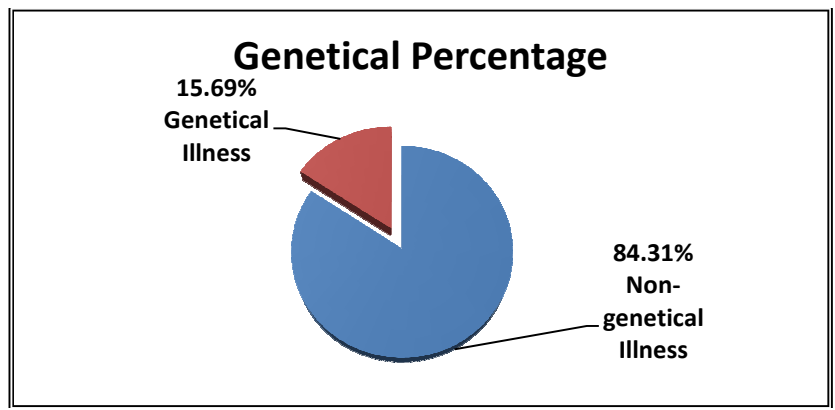

Figure 5: Percentage of parents that have the same illness.

Figure 6 shows most of the patients are from the rural area. Rural area patients contribute 86.27 percent compared to urban area patients with a 13.73 percent contribution. As we know, the rural area in Tawau may suffer from low hygienic surroundings and lack of knowledge on good health. They still do open burning and most of their sources of water are from the nearby rivers or streams. The government should be well aware of this inadequacy in order to prevent any undesired illnesses among the children in the future. On the other hand, Figure 7 shows that, besides environment and genetical factors, the children were infected by pneumonia through several diseases. These diseases had caused the children to be easily infected by pneumonia. Bronchial asthma contributes 11 percent and it's normally caused by a combination of genetic and environmental factors.

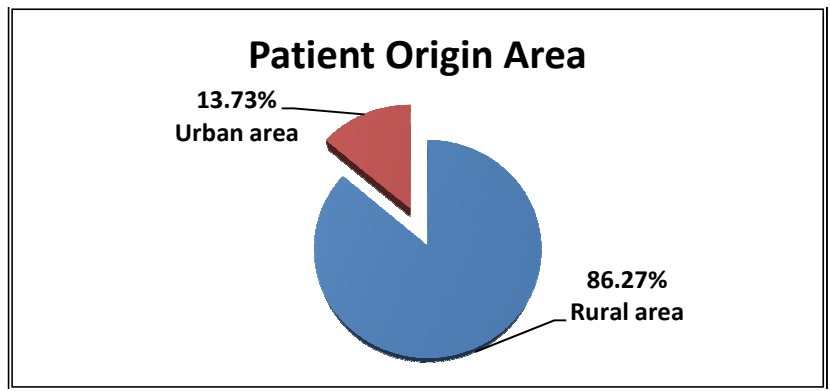

Figure 6: Percentage of patient origin area.

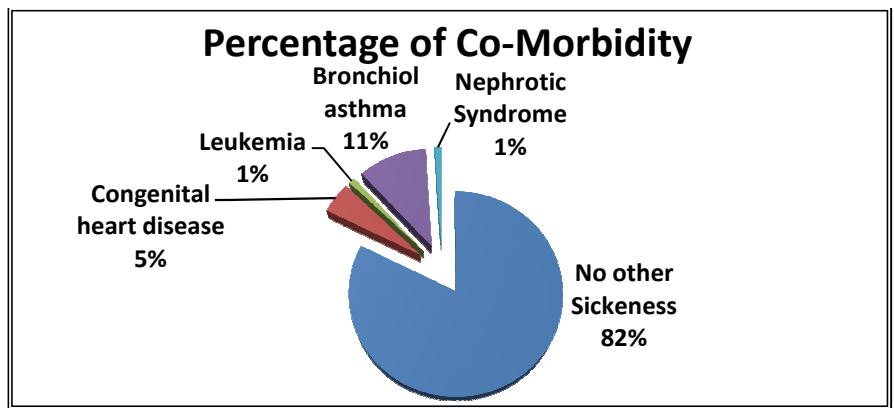

Figure 7: Percentage of co-morbidity.

### 5. CONCLUSION

Pneumonia is one the most dangerous diseases affecting children under 12 years old. Based on the data, we manage to do some statistical analysis and produce some results. The fluctuations of the infection of pneumonia among children are influenced by many factors, but most studies are focusing on hygiene. It has been the main objective of the Malaysian Health Ministry to prevent any disease infection and create a better life for children in Malaysia. The purpose of this article is to incorporate the prior process of the information-transmission before pneumonia infection among children under 12 years old. Few prevention steps can be done in order to stop the spreading of the diseases. We will extend the research in certain part of the Johor region in Malaysia.

**Acknowledgments**
Funding: University Tun Hussein Onn Malaysia, Parit Raja, Batu Pahat, Johor, Malaysia.